# The Periodic Signals of Nova V1674 Herculis (2021)

Joseph Patterson,[1] Marguerite Epstein-Martin,[2] Josie Enenstein,[3] Jonathan Kemp,[4] Richard Sabo,[5] Walt Cooney,[6] Tonny Vanmunster,[7] Pavol Dubovsky,[8] Franz-Josef Hambsch,[9] Gordon Myers,[10] Damien Lemay,[11] Kirill Sokolovsky,[12] Donald Collins,[13] Tut Campbell,[14] George Roberts,[15] Michael Richmond,[16] Stephen Brincat,[17] Joseph Ulowetz,[18] Shawn Dvorak,[19] Tamás Tordai,[20] Sjoerd Dufoer,[21] Andrew Cahaly,[22] Charles Galdies,[23] Bill Goff,[24] Francis P. Wilkin,[25] Enrique de Miguel,[26] and Matt Wood[27]

[1]*Department of Astronomy, Columbia University, 550 West 120th Street, New York, NY 10027, USA, jop@astro.columbia.edu*
[2]*Department of Astronomy, Columbia University, 550 West 120th Street, New York, NY 10027, USA, mae2153@columbia.edu*
[3]*University of Pennsylvania Department of Physics and Astronomy, 209 S 33rd St, Philadelphia, PA 19104, USA, josieenenstein@gmail.com*
[4]*Mittelman Observatory, Middlebury College, Middlebury, VT 05753, USA jkemp@middlebury.edu*
[5]*CBA-Montana, 2336 Trailcrest Drive, Bozeman, MT 59718, richard@theglobal.net*
[6]*Madrona Peak Observatory, 1057 Mickle Creek Rd, Medina, TX 78055, USA, waltc1111@gmail.com*
[7]*CBA-Belgium, Walhostraat 1A, B-3401 Landen, Belgium, tonny.vanmunster@gmail.com*
[8]*Vihorlat Observatory, Humenne, Mierova 4, 06601, Slovakia, var@kozmos.sk*
[9]*CBA-Mol, ROAD Observatory, Oude Bleken 12, B-2400 Mol, Belgium, hambsch@telenet.be*
[10]*CBA-San Mateo, 5 Inverness Way, Hillsborough, CA 94010, USA, gordonmyers@hotmail.com*
[11]*CBA-Quebec, 195 Rang 4 Ouest, St-Anaclet, Quebec G0K 1H0, Canada, damien.lemay@telus.net*
[12]*Center for Data Intensive and Time Domain Astronomy, Department of Physics and Astronomy, Michigan State University, 567 Wilson Road, East Lansing, MI 48824, USA, kirx@kirx.net*
[13]*Warren Williams College, 701 Warren Wilson Road, Swannanoa, NC 28778, USA, dcollins@warren-wilson.edu*
[14]*CBA-Arkansas, 7021 Whispering Pine Road, Harrison, AR 72601, USA, jmontecamp@yahoo.com*
[15]*CBA-Tennessee, 2007 Cedarmont Drive, Franklin, TN 37067, USA, georgeroberts0804@att.net*
[16]*Rochester Institute of Technology, Physics Department, Rochester, NY 14623, USA, mwrsps@rit.edu*
[17]*Flarestar Observatory, San Gwan SGN 3160, Malta, stephenbrincat@gmail.com*
[18]*CBA-Illinois, Northbrook Meadow Observatory, 855 Fair Lane, Northbrook, IL 60062, USA, joe700a@gmail.com*
[19]*CBA-Orlando, Rolling Hills Observatory, 1643 Nightfall Drive, Clermont, FL 34711, USA, sdvorak@rollinghillsobs.org*
[20]*CBA-Hungary, Budapest, Hungary, ttordai01@t-online.hu*
[21]*CBA-DFS, DFS Fregenal de la Sierra, Spain, sdufoer@gmail.com*
[22]*Union College, 807 Union Street, Schenectady, NY 12308, USA, andrewcahaly@outlook.com*
[23]*Institute of Earth Systems, University of Malta, Malta, cgaldies@gmail.com*
[24]*AAVSO, 49 Bay State Road, Cambridge, Massachusetts 02138, USA, astroguyinsc@gmail.com*
[25]*Department of Physics and Astronomy, Union College, 807 Union St, Schenectady, NY 12308, wilkinf@union.edu*
[26]*CBA-Huelva, Universidad de Huelva, E-21070, Huelva, Spain, edmiguel63@gmail.com*
[27]*Texas A&M University, Physics and Astronomy Department, College Station, Texas 77843, USA, matt.wood@tamuc.edu*

## ABSTRACT

We present time-series photometry during eruption of the extremely fast nova V1674 Herculis (Nova Her 2021). The 2021 light curve showed periodic signals at 0.152921(3) d and 501.486(5) s, which we interpret as respectively the orbital and white dwarf spin-periods in the underlying binary. We also detected a sideband signal at the *difference* frequency between these two clocks. During the first 15 days of outburst, the spin-period appears to have increased by 0.014(1)%. This increase probably arose from the sudden loss of high-angular-momentum gas ("the nova explosion") from the rotating, magnetic white dwarf. Both periodic signals appeared remarkably early in the outburst, which we attribute to the extreme speed with which the nova evolved (and became transparent to radiation from the inner binary). After that very fast initial increase of ∼71 ms, the spin-period commenced a steady decrease of ∼160 ms/year — about 100x faster than usually seen in intermediate polars. This is probably due to high accretion torques from very high mass-transfer rates, which might be common when low-mass donor stars are strongly irradiated by a nova outburst.

## 1. INTRODUCTION

V1674 Herculis was an exceptionally bright classical nova which erupted on 12 June 2021. It was first seen at 11th magnitude by S. Ueda (CBET 4976), and its meteoric one-day rise from $V = 20.5$ to 6.3 was reported in detail by Quimby, Shafter, & Corbett (2021). There followed a frenzy of rapid announcements (spectrum, reddening, radio, X-ray, gamma ray, lack of neutrinos, etc.) which confirmed it was "just a nova" (Munari et al. 2021a, b; Wagner et al. 2021; Woodward et al. 2021; Drake et al. 2021). Despite this stain of non-uniqueness, it is still perhaps the most interesting nova of the present young century. Having erupted in Hercules and in June, crossing meridians near local midnight, the star is likely to be the subject of many more studies of the eruption and aftermath. We were among the beguiled, and used the Center for Backyard Astrophysics (CBA, Patterson et al. 2013) worldwide network of small telescopes to carry out time-series photometry from day 3 ($V = 8$) to day 350 ($V = 17$). In this paper we report on that first year of coverage, which comprises ∼1500 hours at a time resolution of ∼60 s.

## 2. THE ORBITAL SIGNAL

Starting on day 4, at $V = 10$, our data revealed an apparent signal with a period near 3.6 hours, as suggested in Figure 1a. In those early days, it was necessary to subtract the quickly declining mean brightness, so we could not maintain continuous tracking of this signal until the rate of decline leveled off around day 16 ($V = 12$). After that, it was possible to time all the minima, and we list them in Table 1. Figure 1b gives an O-C diagram of these timings, relative to the ephemeris:

$$\text{Minimum Light} = \text{HJD } 2{,}459{,}400.637 + 0.152921 \text{ E}. \quad (1)$$

Many of the early light curves showed distinctive secondary minima around phase 0.5. We list some of the more prominent "phase 0.5" dips in Table 1 for reference, but do not show them in the O-C diagram. These features weakened with time. The orbital waveform was reasonably stable after day 30 ($V = 13.5$) and is shown in the lowest frame of Figure 1.

Compared to the well-defined periodicity of primary minima in year one (2021), the second year's orbital timings appear to be systematically late in Figure 1b by ∼0.2 cycles. This might be due in part to the lower accuracy of timings near solar conjunction in January, when the runs are necessarily short. Alternatively, it is also possible that the period is increasing at a very high rate (manifested by *curvature* in the O-C diagram). Future timings, increasing the baseline to two full years, will clarify this.

Such a change is plausible in any large mass-loss-or-transfer event, but known examples are mighty few. One is T Pyx, whose period lengthened by 0.005% in its 2011 outburst (see Figures 6-7 of Patterson et al. 2017). Other good candidates are described by Schaefer (2020) and Patterson et al. (2022).

## 3. THE RAPID PULSATIONS

In the early days of the eruption, rapid periodic signals were reported in the soft X-ray (SX) (Maccarone et al. 2021, Page et al. 2021, Pei et al. 2021, Drake et al. 2021, Page et al.

2022) and optical (Patterson et al. 2021, Schmidt et al. 2021). Their high luminosity (~$10^{38}$ erg/s) and very high pulse fraction (>80%) in the SX suggested a collimated high-$\dot{M}$ flow to a small region on a rapidly rotating and magnetic white dwarf (WD). In our optical coverage, they first appeared around 0.01 mag full amplitude near day 12 of the eruption (at $V = 12$) and steadily grew, reaching ~0.09 mag at day 350 ($V = 17$). Figure 2 shows a sample power spectrum early in the eruption, when "sideband" signals were consistently present, displaced by multiples of $\Omega$ (the orbital frequency) from the main pulse signal at frequency $\omega$. The presence of such sideband signals is a classic signature of an intermediate polar (IP), where the WD rotates prograde with respect to the orbit [for a review of IPs, see Patterson (1994)]. The presence of a highly luminous and highly pulsed X-ray component at the same frequency ($\omega$) leaves little doubt that the signal at 501 s is the true rotation period of an accreting, magnetic WD... and the lower-frequency signals are simply orbital sidebands. Those sidebands disappeared after ~70 days, probably due to a fade in the highly luminous supersoft signal at $\omega$ (which presumably powers the sidebands through reflection off, or absorption by, structures orbiting prograde with frequency $\Omega$ — such as the donor star.

As the star faded from outburst, the pulse ("spin" in our interpretation) period smoothly decreased. The measured phase changes are shown by O-C diagrams in Figure 3. Each is consistent with a fast decline of -160±12 ms/year, although the precise solution will await the end of the 2022 observing season, when it should be possible to determine cycle count across solar conjunction.

As observed by Mroz et al. (2021) and discussed in more detail by Drake et al. (2021), the pre-outburst period was 501.4277(4) s. In our first definite detection, near $V = 12$ and around day 15 of the eruption, the period had suddenly increased to 501.499(3) s. This apparent spin-down of 71(4) ms, seemingly in less than 15 days, is remarkable and unprecedented. Figure 3 shows clearly that the subsequent period *decrease* was smooth, with dP/dt ≈ -160 ms/year.

We can gain some perspective on these numbers by considering period changes in IPs which have *not* suffered a nova eruption while the precise period is being tracked — i.e., all the others.

Pulse-period changes in IPs are generally around 1-2 ms/year, with most of them spinning up (Patterson et al. 2020). While the V1674 Her $\dot{P}$ is ~100x higher than normal, the star is also considerably *brighter* than normal IPs. Assuming that the magnitude 20.5 prenova is a normal IP, V1674 Her during most of 2021-2 (the time of our pulse-period measurements) is 3.5-5 magnitudes brighter than "normal." When converted to luminosity, that implies a factor of 30-100x enhancement. Therefore, it is quite possible that *the anomalously high $\dot{P}$ is a natural result of the very high accretion rate (and therefore high accretion **torque**) in the immediate aftermath of a nova eruption.*

During the one-year time span of the pulse-period measurements, the rate of period change remained within 30% of the mean (-160 ms/year). From mid-2021 to mid-2022 (*viz.*, the mid-point of our period measures during those years), the star faded by 0.9±0.2 mag ($V = 16.1$-$17.0$) in 270 days — a decrease in brightness of 2.3x in 0.74 years.

What period change might we *expect* from such a change in $V$ light? In the theory of disk accretion onto magnetic compact stars (Ghosh & Lamb 1983), and specifically in the "slow rotator" case which is likely applicable to the high-$\dot{M}$ phases like a young postnova, the rate of rotation period change should scale as $(\dot{M})^{3/7}$. Thus a 2.3x decline in accretion rate might be expected to cause a drop in $\dot{P}$ by a factor of $(2.3)^{3/7} = 1.4$. But since we have not yet amassed a time baseline sufficient to count spin-period cycles uniquely between years, and since we do not know how much non-accretion light (the shell and/or donor) is present, we do not yet have a test of high precision. A few years of pulse timings may give us one.

## 4. THE PULSED X-RAYS

The observed properties of the optical 501 s signal (short period, high coherence, spin-up, presence of $\omega$ - $\Omega$ sidebands, etc.) make it clear that this is a classic IP. However, the properties of the SX counterpart (same period, $\sim$90% pulse fraction, near-Eddington luminosity at 50-100 eV energies) add substantially to the picture, and prove that there is a sustained, very high accretion rate onto the magnetic pole. This is true whether the energy source is prompt H $\to$ He burning (6 MeV/nucleon) or free-fall accretion onto a massive WD (0.5 MeV/nucleon for $M_1 = 1.2$ $M_O$). In either case, the energy is pulsed at the spin-period, and probably thermalized on the WD's surface at the magnetic poles to produce the 50-100 eV component observed by the SX telescopes.

In the more common theory of accretion onto IPs, the infall energy is released largely in a shock well above the magnetic poles, producing a pulsed luminous component ($10^{33}$ erg/s) in very hard X-rays (20-100 keV). That may be present in V1674 Her, too. But for the much larger accretion rates suggested here by high SX luminosities and large period changes, direct deposition of most infall energy in the photosphere seems more likely.

The optical measures of $P_{spin}$ are accurate to $\sim$0.003 s. Because each SX observation is brief (a few days), the published X-ray period measurements are much less accurate (0.25 s). Still, they all agree within that error, so it is likely that they measure essentially the same thing. Whether the *phases* (e.g., of maximum light) agree is still not known from any of the published work, and will remain so until an actual SX ephemeris is available. This is especially important early in the outburst, because it will test where in the binary the sideband ($\omega$ - $\Omega$) pulses originate.

## 5. THE ORBITAL LIGHT CURVE, REVISITED

In a recent study of supersoft binaries and recurrent novae of particularly high quiescent luminosity ($M_V$ = +3 or brighter), we found a characteristic pattern: a double-humped orbital light curve and a rapidly increasing orbital period (Patterson et al. 2022). We ascribed the three defining characteristics (high luminosity, intrinsically blue color, and increasing $P_{orb}$) to one cause: a particularly high and prolonged accretion rate ($\sim 10^{-7}$ Mo/year). The idea is that high accretion onto the WD irradiates the donor star with a greater flux, and the donor responds by sending a greater $\dot{M}$ over to the WD. That high $\dot{M}$ from the low-mass member of the binary increases $P_{orb}$, assuming that loss of angular momentum does not counteract it.

If the accreted matter is converted to luminosity via H $\to$ He nuclear reactions, the efficiency can be quite high (King & van Teeseling 1998, Knigge et al. 2000), powering the more-or-less-permanent supersoft sources. *Gravitational* luminosity is less efficient, inherently making this "bootstrap" process less efficient. In the latter case, the mechanism probably requires tapping some of the energy left over from the latest nova outburst. That energy will presumably expire in a few years or decades.

In the meantime, the (hypothetically and transiently) luminous donor star will not necessarily reveal itself in spectra, if the energy deposited by its more luminous neighbor arrives in the donor's "reversing layer," where the absorption lines are formed. External energy deposited there, top-down, can possibly destroy the temperature gradient necessary for absorption lines.

This hypothesis merits further theoretical study. In part, it appeals to us because our study of post-nova orbital light curves suggests that *most* are double-humped — consistent with eclipses of two luminous and large regions (disk + donor?) separated by exactly 0.5 in orbital phase. Such light curves, particularly those displaying secondary minima, are rare among cataclysmic variables which have not experienced a recent nova outburst.

With its lingering high soft X-ray luminosity and possible large increase in orbital period, V1674 Her may test our understanding of these matters.

## 6. SUMMARY

1. We trace the changes of the 501 s WD rotation period in V1674 Her during 2021 and 2022. After it could first be measured around day 12 of the eruption, the period decreased at a rate of 160±12 ms/year, even as the star faded by 4 mag (13-17) in this first-ever tracing of the changes in WD rotation during the first year of life as a post-nova.

2. The initial measure of the spin-period (around day 15) showed that the period had lengthened suddenly by 71±5 ms relative to the value deduced by Mroz et al. (2021) in quiescent-state data from the Zwicky Transient Facility (ZTF). This change likely arose from angular momentum loss in the ejecta ("magnetic braking"?), as also discussed by Drake et al. (2021).

3. The eclipse timings showed some evidence for orbital period increase — a common trait of compact binaries with a very luminous SX component. This is suggested by Figure 1(b), and should be investigated in future years. It seems unlikely that the *pre*-eruption orbital period can now be learned; but with the current $P_{orb}$ now known to 6 significant figures, it is worth a try (from archival all-sky data).

4. In a recent paper on spin-period changes in IPs, we speculated that the observed strong preference of these stars to show spin-up (rather than spin-down) might be due to the after-effects of an ancient nova eruption (the **"Bossa Nova"** theory: Gorme 1963, Patterson et al. 2020). If so, the spin-up rate in V1674 Her should decline slowly, probably on a timescale of centuries or longer.

5. Finally, we were amazed that all the basic features of the orbital light curve were readily visible, and on their orbital schedule, as early as day 4 (10.5 magnitudes above quiescence). This is yet another feature unprecedented in the history of nova studies. Perhaps it is mainly a consequence of (observed) high ejection velocity and (hypothesized) low ejected mass; that combination could "thin out" the ejecta pretty quickly. Alternatively, the ejecta could have been very anisotropic, mostly (and happily) avoiding our precise line of sight to the inner binary. The floor is certainly open to other suggestions.

## 7. ACKNOWLEDGMENTS

This research was supported in part by grants from the National Science Foundation (NSF AST-1908582) and NASA (HST-GO-15454.002-A). The work also includes observations obtained with the Mittelman Observatories 0.5 m telescope at New Mexico Skies in Mayhill, New Mexico. Additionally, we would like to thank the Mittelman Family Foundation for its generous support of and substantial impact on astronomy at Middlebury College. Finally, we thank the Williams Family Foundation for their support.

Table 1. Measured times (HJD - 2,450,000) of minimum light in the 3.67 hour variation.

| | | | | |
|---|---|---|---|---|
| 9382.666* | 9401.870 | 9452.4763 | 9490.4017 | 9662.6099 |
| 9382.820* | 9405.686 | 9455.3788 | 9500.3437 | 9690.5950 |
| 9394.659 | 9405.839 | 9459.3538 | 9511.6577 | 9697.6244 |
| 9394.811 | 9416.5377 | 9460.4277 | 9515.4813 | 9716.5905 |
| 9400.633 | 9424.7932 | 9465.6265 | 9517.4711 | 9740.7620 |
| 9400.711 | 9430.6081 | 9470.3681 | 9519.6056 | 9745.4975 |
| 9400.784 | 9440.5451 | 9476.4827 | 9524.3458 | |
| 9401.387 | 9444.3684 | 9480.3047 | 9527.4045 | |
| 9401.548 | 9448.3465 | 9483.6695 | 9531.3812 | |

*These early timings, while reliable and accurate, are not necessarily continuous with the family of later timings we consider "reliably orbital" in origin.

FIGURE CAPTIONS

Figure 1. (a) Light curve on JD 9382 ($V = 10$) after subtracting the linear decline. The apparent dips seem to correspond to phases near 0.0 and 0.5 in our orbital ephemeris. At a remarkable 7 magnitudes above the 2022 level, the light curve looks basically the same!
(b) O-C diagram of the times of minimum light in the orbital signal, relative to the test ephemeris HJD 2,459,400.636 + 0.152921 E. Omitted are timings of "secondary minima," some of which are included in Table 1.
(c) The average orbital waveform after the first 30 days of outburst.

Figure 2. Power spectrum of a typical 14-day segment of light curve early in the outburst, showing the powerful rotational signal at $\omega$, and orbital sidebands at $\omega - \Omega$ and $\omega - 2\Omega$. The sidebands disappeared after $\sim$60 days.

Figure 3. O-C diagram of the times and spin-period maximum light during 2021 and 2022. The fitted parabolas are each within 20% of -160 ms/year.

FIGURE 1

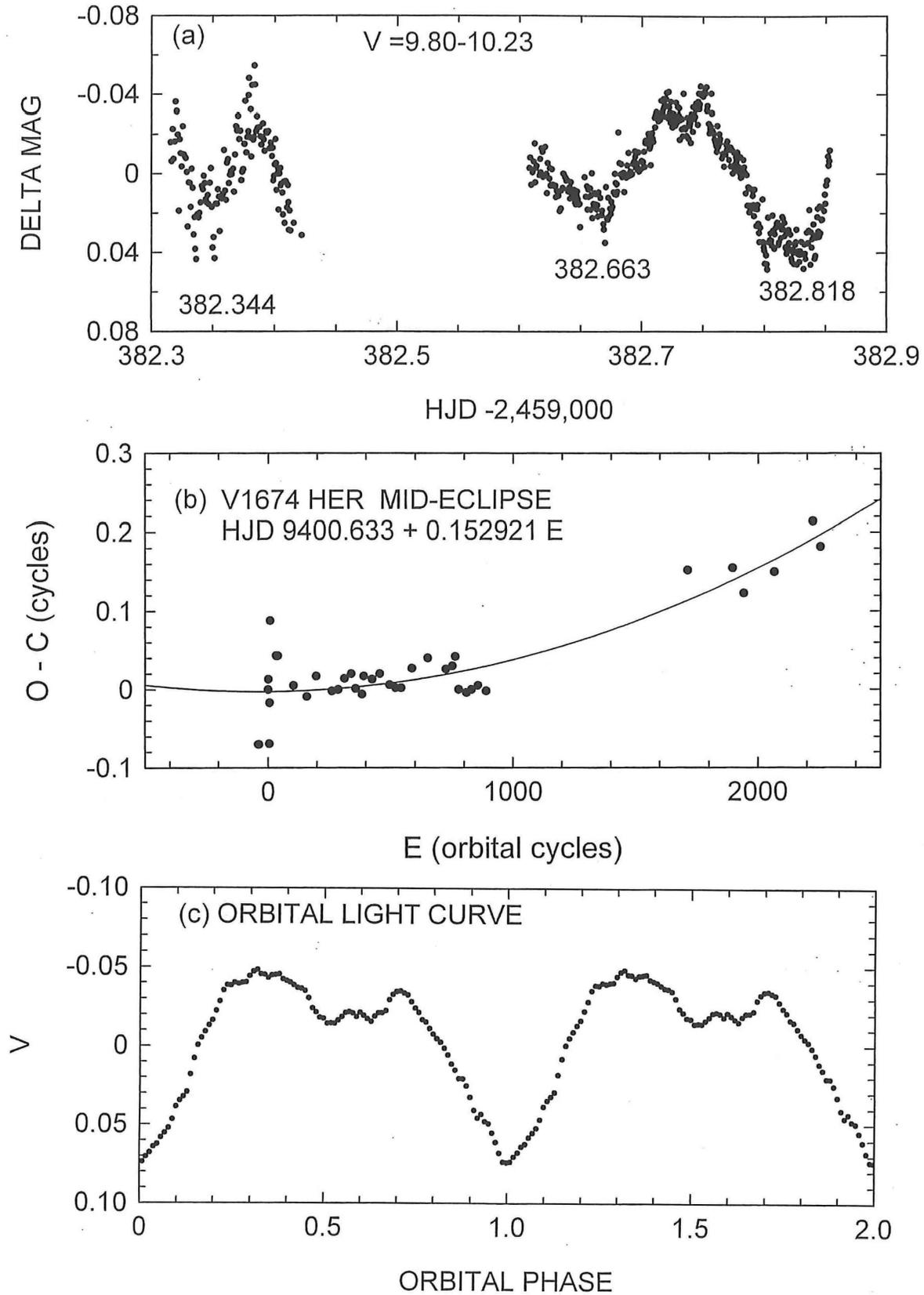

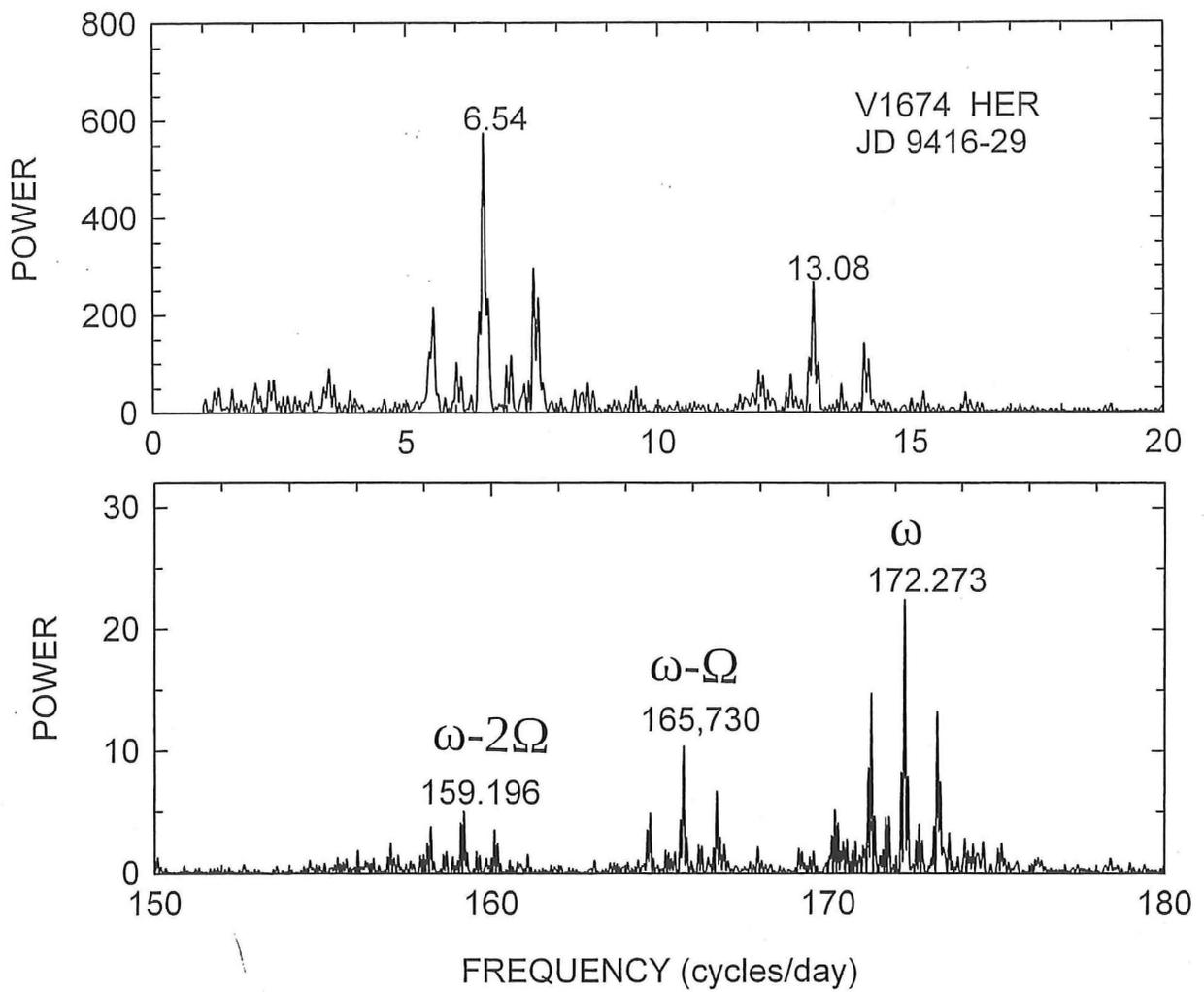

FIGURE 2

FIGURE 3

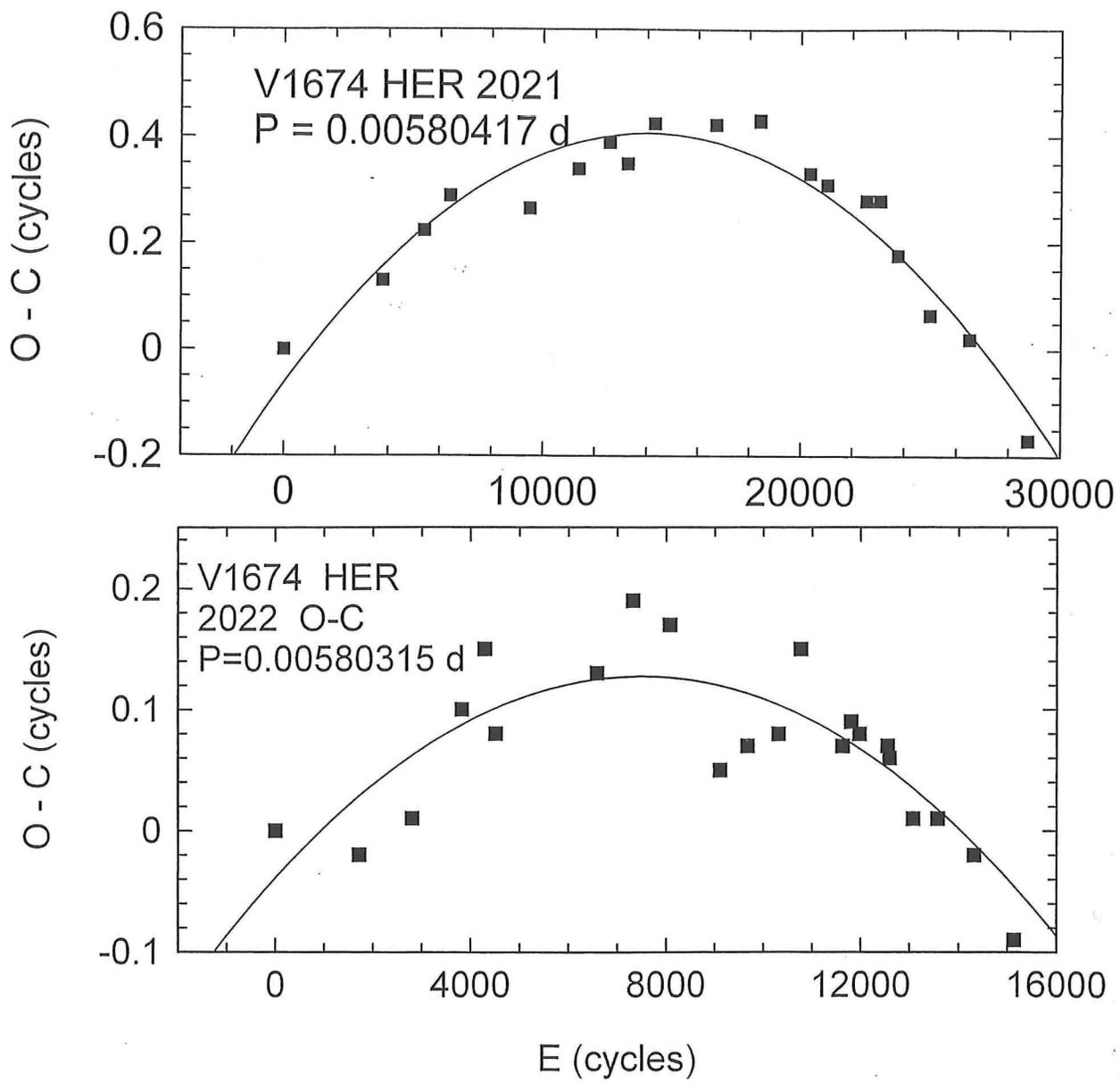